\documentclass[a4paper]{article}

\usepackage[a4paper]{geometry}

\usepackage[utf8]{inputenc}
\usepackage[english]{babel}
\usepackage{amsmath}
\usepackage{amsfonts}
\usepackage{amssymb}
\usepackage{calc}
\usepackage{soul} 

\usepackage{siunitx}
\DeclareSIUnit\fm{\femto\meter}
\DeclareSIUnit\nb{\nano\barn}
\DeclareSIUnit\ly{\text{ly}}
\DeclareSIUnit\mias{\micro\text{as}}
\DeclareSIUnit\msun{\text{\ensuremath{M_\odot}}}
\DeclareSIUnit\lsun{\text{\ensuremath{L_\odot}}}

\usepackage[version=4]{mhchem}

\usepackage[comma,numbers,sort&compress]{natbib}

\usepackage{graphicx}

\usepackage{hyperref}
\hypersetup{
    pdftitle={Black Holes and High Energy Physics: From Astrophysics to Large Extra Dimensions},
    pdfauthor={Michael Florian Wondrak},
    pdfstartview={FitH}
}

\author{Michael Florian Wondrak$^{a,b,}$\thanks{E-mail: \texttt{wondrak@fias.uni-frankfurt.de}} %
\and Marcus Bleicher$^{a,b,}$\thanks{E-mail: \texttt{bleicher@fias.uni-frankfurt.de}}%
\and Piero Nicolini$^{a,b,}$\thanks{E-mail: \texttt{nicolini@fias.uni-frankfurt.de}}\\[1ex]
\small $^a$ Frankfurt Institute for Advanced Studies (FIAS)\\[-0.5ex]
\small Ruth-Moufang-Str.~1, 60438 Frankfurt am Main, Germany\\[1ex]
\small $^b$ Institut f\"{u}r Theoretische Physik, Johann Wolfgang Goethe-Universit\"{a}t Frankfurt\\[-0.5ex]
\small Max-von-Laue-Str.~1, 60438 Frankfurt am Main, Germany}
\date{}
\title{Black Holes and High Energy Physics:\\%
From Astrophysics to Large Extra Dimensions}

\newcommand{\dd}{\ensuremath{\text{d}}}
\newcommand{\lambdabarnew}{{\;\mkern 0.75mu\mathchar '26\mkern -9.75mu\lambda}}
\newcommand{\GBareHigher}{\ensuremath{G_D}}
\newcommand{\horizRad}{\ensuremath{r_\text{h}}}
\newcommand{\Rc}{\ensuremath{R_\text{c}}}
\begin{document}
\maketitle

\begin{abstract}
\noindent Up to now, Einstein's general theory of relativity has passed all experimental tests. But yet we know that it is not a fundamental theory and that it is incompatible with quantum theory. While several extended and improved gravitational theories on classical and quantum level are nowadays available, it is a great challenge to find experimental setups to check their signatures.

We discuss recent developments in direct observation of black holes comprising gravitational waves from black hole mergers, radio interferometry images of black hole shadows, and Hawking radiation of black holes in particle accelerators. These investigations cover the full black hole mass range from microscopic to stellar and supermassive black holes. We comment on the associated strong-field tests of Einstein's general theory of relativity and implications for quantum gravity. 

Special emphasis lies upon the physics of large extra dimensions and black hole evaporation, the existence of a minimal black hole mass, and the cross sections of higher-dimensional black holes. We complete this short review with the latest experimental constraints at the Large Hadron Collider.
\end{abstract}

\begin{quote}
\begin{flushright}   
``Theoretical physicists read in the book of nature \\%
--- but experimentalists turn its pages.''\\[1ex]%
\em Walter Greiner (1935--2016)
\end{flushright}    

\end{quote}

\section{Introduction}
In these days, gravitational physics and astrophysics enter a new era: Black holes, which belong to the most astounding objects resulting from Einstein's general theory of relativity, come into reach of direct observation. 
Previous experimental studies of black holes have only been possible indirectly, \textit{e.g.}, by observing the matter-donating companion star in an X-ray binary system \cite{Bolton1972,WebsterM1972}. Using gravitational wave experiments and very-long-baseline interferometry, black holes can now be studied directly \cite{VirgoLIGO2016a,FalckeMA1999}. 
Although these astronomically observable black holes have masses from few to billions of solar masses ($\si{\msun}$), there is the speculative possibility of much smaller black holes with masses in the $\si{\TeV}$ range, \textit{i.e.}, potentially being produced in high-energy particle collisions \cite{ADM98,BaF99,DiL01,GiT02,CMS2017}. Since improved experimental techniques have been developed, one can hope to 
distinguish in the near future between black holes resulting from Einstein's theory and those from alternative theories, including the case of completely different massive objects \cite{HessMG2010}. 

In section \ref{sec:BHs}, we describe the possibility of testing the concept of general relativistic black holes and alternative ideas for new astronomical observations. We go on to review the idea of large extra dimensions with special emphasis on the production of black holes in accelerators and report on the experimental status in section \ref{sec:LXD}. In section \ref{sec:over&out}, we draw our conclusions and give an outlook.

\section{Astrophysical Black Holes}
\label{sec:BHs}
The concept of black holes goes back to the 18th century.  Based on Descartes' and Newton's corpuscular model, 
Mitchell suggested, in 1784,  that even light could not escape from a massive object if its gravitational force is strong enough. 
The first black hole solution to Einstein's field equations has been derived by Schwarzschild in 1916, but it has been recognized as such by Finkelstein only in 1958. In the late 1960's, Hawking and Penrose showed the role of black holes as generic end-stage solutions for gravitational collapses within the Einstein's theory of relativity \cite{HawkingP1969}. 

On the observational side, however, the situations is less clear. The closest available hints  are indirect. Observations suggest the existence of compact massive objects whose spatial extensions are too small to be composed of several ordinary astrophysical objects or neutron stars. For instance, studies of the trajectories of the closest stars to the galactic center showed that there is a total mass of $\SI{4.3e6}{\msun}$ confined in a region of $\SI{4.4e10}{\meter}$ radius at the center of our Milky Way \cite{GillessenETAGMO2008} (cf.~\cite{GhezEtAl2008,SchodelME2009}). 
Nowadays, however, experimental techniques, such as gravitational wave detectors and radio interferometers,  have drastically improved and may allow for a direct study of black hole properties.

The first gravitational wave signal has been detected in September 2015 by the LIGO and Virgo collaborations \cite{VirgoLIGO2016a}. In the meantime, two further signals and one candidate signal with decreased signal-to-noise ratio have been recorded \cite{VirgoLIGO2016b,VirgoLIGO2016c,VIRGOLIGO2017}. 
The analyses of these events strongly support black hole mergers as sources while they exclude the case of neutron star mergers. According to the simulations, the mergers took place in a distance between $\SI{1.3e9}{\ly}$ and $\SI{2.9e9}{\ly}$ and the participating black hole masses were in the range between $\SI{8}{\msun}$ and $\SI{36}{\msun}$.

The detected signals perfectly agree with predictions based on black holes in Einstein's general relativity up to $5\sigma$ \cite{VirgoLIGO2016c}. Especially the attenuation of the signal in the ring-down phase (the phase after the actual merger when the final black hole settles) indicates a photon ring which is only present if the object is compact enough \cite{CardosoFN2016}.
However, there is room for other objects to generate such a gravitational wave signal. For instance, according to \cite{KonoplyaZ2016,DeLaurentisPLAA2016}, the precision is yet not sufficient to distinguish between general relativistic black holes and those of alternative theories of gravity. 

In contrast to gravitational wave findings of stellar-mass black holes, the very-long-baseline interferometry (VLBI)  concentrates on imaging of the supermassive species. The latter possess the largest spatial extensions and distinctive accretion disks so that they subtend the largest angles for an observer on Earth. The synchronized phase-sensitive recordings of many radio telescopes on several continents and subsequent data correlation resemble a radio telescope whose dish size lies in the order of Earth's diameter. The Event Horizon Telescope collaboration (EHT) reaches a resolution of $\SI{35}{\mias}$ which corresponds to the apparent size of a tennis ball on the Moon as seen from Earth. The aim is to record the black hole shadow, \textit{i.e.}, the silhouette of the black hole in front of its background including gravitational lensing effects, and to resolve the origin of jets. The EHT had a promising measurement campaign in April 2017. The data analysis is expected to release the first results in 2018.

The primary focus of the EHT lies upon Sagittarius A* (Sgr A*), the black hole in the center of our Milky Way. As a radio source, its luminosity on the longest wavelengths (namely frequencies up to $\SI{1}{\tera\hertz}$) amounts to $\sim \SI{100}{\lsun},$ where $\si{\lsun}$ is the Sun's luminosity over the whole spectrum. Even though small on the scale of supermassive black holes, this brightness resulted in the discovery of Sgr A* by radio interferometry in 1974. Furthermore, at a distance of $\SI{2.6e4}{\ly}$, it is the closest supermassive black hole having a mass of $\SI{4.3e6}{\msun}$. The corresponding angle, $\SI{53}{\mias}$,  is the largest among all supermassive black holes and allows for radio observation.

The second object of interest is the black hole in the center of the giant elliptical galaxy M87. With a mass of $\SI{6e9}{\msun}$, it is the most massive black hole known. M87 lies in the center of the Virgo galaxy cluster. Even though its distance from Earth is $\SI{5e7}{\ly}$, it represents the center of the Virgo supercluster which the local group, including the Milky Way, belongs to. Because of its size, the black hole's angular diameter, $\SI{22}{\mias}$, is of the same order as the EHT resolution. M87 is of special interest due to its strongly collimated jet of high-energy particles which extends up to $\SI{2.5e5}{\ly}$ from the central black hole.

The investigation of the shadow serves as an experimental testbed for gravitational theories. For Einstein's gravity black holes, the shape depends on the mass and angular momentum only. However, some alternative theories of gravity also lead to modifications of the shape, \textit{e.g.}, \cite{Moffat2015,GrenzebachPL2015}.
One such theory is pseudo-complex general relativity (pc-GR) proposed by Hess and Greiner in order to prevent 
the occurrence of singularities \cite{HessG2009,HessMG2010,CasparSHSG2012,HessSG2016}. Their ansatz algebraically generalizes spacetime coordinates to pseudo-complex, or hypercomplex, numbers which leads to new metric degrees of freedom. In contrast to general relativity, a new energy-momentum contribution arises which is similar to dark energy. Due to this repulsive pressure, a gravitational collapse finally comes to rest and produces a compact object of finite size, a so-called gray star, instead of a black hole. The predicted shadow of a gray star in pc-GR differs from the one of black holes in ordinary general relativity \cite{SchonenbachCHBMSG2012,SchonenbachCHBMSG2013}. Thus this theory could be tested in the future when the EHT gains further resolution.

There is some hope that new, unexpected effects occur in the strong-field observations which have a quantum origin, \textit{e.g.}, probable echoes in gravitational wave signals \cite{AbediHA2016,AbediHA2017}, and that those hint at a more fundamental theory of  physical interactions \cite{GBUP}. Several reasons indicate that both the Standard Model of particle physics as well as Einstein's general relativity are not fundamental. Most obviously, we do not know a consistent theory which unifies all interactions. The Standard Model neither explains the non-zero neutrino masses nor the background of the external parameters. General relativity at classical level is plagued by long standing problems such as curvature singularities, the unknown nature of dark matter, and the cosmological constant problems (\textit{e.g.}, \cite{AfshordiCDG2007,Afshordi2008,Wondrak2017}). Moreover, it lacks a direct quantization  due to its non-renormalizable character.

At a recent conference%
\footnote{3rd Karl Schwarzschild Meeting on Gravitational Physics and the Gauge/Gravity Correspondence, Frankfurt am Main, July 2017} more than 80\% of the participants were convinced that within the next decade there will be observational evidence for quantum gravity. It could be possible to gain this experimental access by a different approach. On the other side of the black hole mass spectrum, microscopic black holes might be produced in particle accelerators, due to increased energy capabilities, \textit{e.g.}, at the Large Hadron Collider as we will see in the next section \cite{Lan02,Cav03,Kan04,Hos04,CaS06,Ble07,Win07,Nic09,BlN10,BlN14,Cal10a,Par12,KaW15,WondrakNB2016}.

\section{Large Extra Dimensions}
\label{sec:LXD}
For the discussion of gravity at small scales, we follow the lines of \cite{WondrakNB2016}. The difference between the Standard Model interactions and general relativity manifests itself impressively in their dissimilar coupling strengths. While the electromagnetic, weak and strong forces differ by just 6 orders of magnitude,
the gravitational interaction falls apart by further 33 orders. The ratio between the Fermi constant, $G_\text{F}$, and the gravitational constant, $G_\text{N}$, amounts to
\begin{equation}
\frac{G_\text{F}}{G_\text{N}} 
\approx \num{1.7e33}.
\end{equation}
The weakness of gravity compared to the other forces is called the hierarchy problem.

The gravitational constant is related to the fundamental Planck mass, $M_\text{Pl}$, by
\begin{equation}
G_\text{N} = \frac{1}{M_\text{Pl}^2},
\end{equation}
where $M_\text{Pl} \approx \SI{1.2e16}{\TeV} \approx \SI{2.2e-8}{\kg}$.
Since $G_\text{N}$ couples to two masses, $m_1$ and $m_2$, the term $G_\text{N}\,m_1\,m_2$ has to be of the order of 1 for reaching a  strength comparable to the other interactions. Thus, quantum gravity is supposed to set in at the Planck scale.

The Standard Model is an excellent description of particle physics corroborated by  experimental evidence --- up to the LHC energy scale of $\sqrt{s} = \SI{14}{\TeV} = \SI{2.2e-6}{\J}$. The gap of 15 orders of magnitude up to the Planck scale cannot be accessed experimentally yet. Since we do not know the physics in these energy regimes, one could simply assume that no new particles or physical effects emerge, which is named the big dessert approximation. Another idea to tackle the hierarchy problem is to find a mechanism which reduces the fundamental gravitational energy scale. This is realized in the extra dimension ansatz proposed by Arkani-Hamed, Dimopoulos, and Dvali together with Antoniadis, the so-called ADD model \cite{Arkani-HamedDD1998,AntoniadisADD1998,Arkani-HamedDD1999}.

The concept of introducing a higher dimensional spacetime goes back to Kaluza and Klein who tried to unify gravity and classical electrodynamics by adding an additional spatial dimension to form a 5-dimensional spacetime \cite{Kaluza1921,Klein1926a,Klein1926b}. The Maxwell equations directly emerge from the higher dimensional Einstein field equations. Furthermore, the quantization of electric charge automatically appears by compactifying the extra dimension.
Apart from the ADD model, extra dimensions are also employed in (9+1)-dimensional super string theory and (10+1)-dimensional M-theory.

ADD proposed that our familiar (3+1)-dimensional world is just a slice, a so-called D$3$-brane, which is embedded in a higher-dimensional bulk spacetime having $k$ additional spatial dimensions.
While Standard Model fields are confined to the brane, the extra dimensions are only accessible for gravitational fields. In this way, non-gravitational physics is the same while gravitational physics on small scales becomes much stronger. The strength of gravity can be understood by the new profile of the extra-dimensional Newton's law. However, in order to comply with short scale precision tests of gravity, which confirm the inverse square law down to $\SI{56}{\micro\meter}$  \cite{AdelbergerGHHS2009}, the additional dimensions have to be compactified. 

The theory introduces two new degrees of freedom: the number of spatial extra dimensions, $k$, and the new fundamental gravitational energy scale, 
\begin{equation}
M_D \equiv {\left( \frac{1}{8\pi}\,\frac{1}{\GBareHigher} \right)}^{1/\left(2+k\right)},
\end{equation}
where $\GBareHigher$ is the gravitational constant in $D$=((3+$k$)+1) dimensions.
Relating $\GBareHigher$ to the ordinary 4-dimensional gravitational constant $G_\text{N}$ determines the compactification radius, $\Rc$, of the extra dimensions,
\begin{equation}
\Rc 
= \frac{1}{2\pi}\,{\left(\frac{M_\text{Pl}^2}{8\pi\,M_D^{2+k}}\right)}^{1/k} 
\sim 10^{\frac{32}{k}-19}\si{\meter},
\label{eq:CompactRadius}
\end{equation}
where we assumed $M_D \sim \SI{1}{\TeV}$ in the last step. In this case, consistency with experiments requires $k \geq 3$ and $\Rc$ follows to be in the range $\left[\si{\fm},\,\si{\nano\meter}\right]$.

Analogously to the static, uncharged Schwarzschild black hole, we can find a black hole solution to Einstein's field equations in the higher-dimensional spacetime%
\footnote{For a discussion about higher dimensional gravity field equations cf.~\cite{Dadhich2015}. For the repercussions of higher dimensional scenarios in gravitational waves cf.~\cite{AndriotG2017}.} %
\cite{Tangherlini1963,Myers1986}. The line element is given by
\begin{equation}
\dd s^2
= -\left(1-{\left(\frac{\horizRad}{r}\right)}^{1+k}\right) \, \dd t^2
  +\frac{1}{\left(1-{\left(\dfrac{\horizRad}{r}\right)}^{1+k}\right)} \, \dd r^2
  +r^2 \, \dd \Omega_{2+k}^2
\end{equation}
where $\dd \Omega_{2+k}^2$ is the usual angular line element in $2+k$ dimensions. The event horizon is located at
\begin{equation}
\horizRad
={\left(\frac{\Gamma\!\left(\frac{3+k}{2}\right)\,M_\text{BH}}{\left(2+k\right)\,\pi^{\left(3+k\right)/2}\,M_D^{2+k}}\right)}^{1/\left(1+k\right)}.
\label{eq:hievho}
\end{equation}
For microscopic black holes with masses $M_D \lesssim M_\text{BH} \ll M_\text{Pl}$, the event horizon is much smaller than the compactification radius $\Rc$. This ensures the insensitivity of the black hole to the compactification of dimensions.

A fundamental concept in quantum physics is the delocalization of particles. The typical length scale of the probability distribution 
is given by the reduced Compton wavelength,
\begin{equation}
\lambdabarnew_\mathrm{C} 
= \frac{1}{m}.
\end{equation}
Starting from this principle we naturally derive the existence of a lower mass limit for black holes. The mass of a black hole defines its horizon radius \eqref{eq:hievho}. In order not to leak out, the black hole has to be self-contained within its reduced Compton wavelength. If both characteristic scales agree, a black hole has the minimal mass $M^\text{min}_\text{BH}$ which amounts to
\begin{equation}
M^\text{min}_\text{BH} 
\sim M_D \,
\sim {\left(\dfrac{l_\text{Pl}}{\Rc}\right)}^{k/\left(2+k\right)}\,M_\text{Pl}.
\label{eq:BH_min_approx}
\end{equation}
where $l_\text{Pl} \approx \SI{1.62e-35}{\meter}$ is the ordinary 4-dimensional Planck length. Of course, this argumentation assumes that the concepts of the Compton wavelength and of the classical horizon hold up to the fundamental energy scale $M_D$. The actual minimal mass will be determined by a theory of quantum gravity \cite{GRW02,AuS13,DvG10,DFG11,DGG11,DvG12,DvG13b,DvG13,DvG14,DGI15,CaS14,CMS14,CMN15b,CMOr12,MuN12,
IMN13,FKN16}.

If $M_D$ lies in the range of $\si{\TeV}$, microscopic black holes could be produced in particle accelerators \cite{ADM98,BaF99,DiL01,GiT02}. The LHC center-of-mass energy for pp collisions lies at $\sqrt{s} = \SI{14}{\TeV} = \SI{2.2e-6}{\J}$ at a design luminosity of $L = \SI{e10}{\per\barn\per\s}$ which has already been exceeded in the last year \cite{CERN2016}. In head-on collisions of $\ce{^{208}_{82}Pb}$, the mean center-of-mass energy per nucleon pair amounts to $\SI{2.76}{\TeV}$.
If two partons with normalized initial longitudinal momentum $x_1$ and $x_2$, respectively, collide at a center-of-mass energy of 
$\sqrt{\hat{s}} = \sqrt{x_1\,x_2\,s} = M_\mathrm{BH}$, they might turn into a black hole of that mass.

A natural ansatz for the production cross section at parton level is the black disk approximation \cite{Thorne1972} which uses the geometric cross section of the event horizon,
\begin{equation}
\hat{\sigma}\!\left(M_\mathrm{BH},\,k\right) 
= \pi\, \horizRad^2\!\left(M_\mathrm{BH}\right).
\end{equation}
However, this ansatz allows the formation of arbitrarily small black holes if the impact parameter in the parton-parton collision is just small enough. To cure this deficit, an improved cross section has been proposed in \cite{MureikaNS2012},
\begin{equation}
\hat{\sigma}\!\left(M_\mathrm{BH},\,k\right)
= \pi l^2 \, \Gamma\!\left(-1;\,\frac{l^2}{\horizRad^2\!\left(M_\mathrm{BH}\right)}\right) \,
  \theta_l\!\left(M_\mathrm{BH}-M^\mathrm{min}_\mathrm{BH}\right).
\end{equation}
It takes into consideration that at the energy level of the fundamental scale $M_D$, a finite resolution of spacetime of size $l$ emerges as suggested by several theories, \textit{e.g.}, \cite{Hos12,SNB12,NSS06b,BoR00,MMN11,Mod06,Nic12}. $\Gamma\!\left(\alpha;\,x\right)$ denotes the upper incomplete gamma function
\begin{equation}
\Gamma\!\left(\alpha;\,x\right)
\equiv \int_x^\infty\!\dd t \; t^{\alpha-1}\,e^{-t},
\end{equation}
and $\theta_l\!\left(x\right)$ the modified Heaviside step function
\begin{equation}
\theta_l\!\left(x\right)
\equiv \frac{1}{{\left(4\pi\,l^2\right)}^{\frac{1}{2}}} \, 
       \int_{-\infty}^x \!\dd y\; \mathrm{e}^{-\frac{y^2}{4 l^2}}.
\end{equation}

The differential cross section of black hole production with respect to the black hole mass $M_\text{BH}$ depends on the parton distribution functions $f_i\!\left(x_j,\,Q^2\right)$ of parton type $i$ in hadron $j$. It turns out to be
\begin{equation}
\frac{\dd \sigma}{\dd M_\text{BH}}
= \sum_{a,\,b} \int_{0}^{1}\!\dd x_1 \; \frac{2M_\text{BH}}{x_1\,s}\,\,
  f_a\!\left(x_1,\,M_\text{BH}^2\right)\,\,f_b\!\left(\frac{M_\text{BH}^2}{x_1\,s},\,M_\text{BH}^2\right)\,\hat{\sigma}\!\left(M_\text{BH},\,k\right).
\end{equation}
For instance, assuming an integral cross section of $\SI{10}{\nb}$, we would expect $\num{e2}$ black holes per second to be produced in pp collisions at the LHC design luminosity \cite{BleicherHHS2002,HossenfelderHBS2002}.

Once a microscopic black hole is produced, it suffers a loss of mass due to the emission of Hawking radiation. The evaporation closely follows a blackbody spectrum where the  temperature is given by the Hawking temperature%
\footnote{Due to the gravitational potential, part of the emitted radiation is reflected so that an observer at spatial infinity actually perceives a graybody modification depending on the particle spin, angular momentum, and frequency.} \cite{Haw75}
\begin{equation}
T_\mathrm{BH}
= \frac{\hbar c\,\kappa}{2\pi\,k_\text{B}}.
\end{equation}
Here $\kappa$ denotes the black hole's surface gravity which grows the faster, the smaller the black hole becomes. This implies that the evaporation rate becomes increasingly large during the evaporation, with the smallest black holes producing the largest radiation densities.

The evaporation process of microscopic black holes can be classified in several stages: the balding phase, the spin-down phase, and the Schwarzschild phase. In the first one the black hole radiates away its electric and other gauge field charges and stabilizes towards a fast rotating state by emission of gravitational waves. In the second phase, its angular velocity slows down due to Hawking and Unruh-Starobinskii radiation. When the hole configuration switches to a static Schwarzschild one, the radiation is emitted with spherical symmetry and one speaks of third phase.

The stages above are valid for black holes with masses (far) above the fundamental mass scale, $M_D$, and based on semiclassical analysis. Close to $M_D$, the description of so-called quantum black holes critically depends on the full theory of quantum gravity which is not yet available. However, several effective models agree in predicting a SCRAM%
\footnote{This term has been adopted in \cite{Nic09} from the emergency shutdown of nuclear reactors.
It originates from the Manhattan Project at Chicago Pile-1 in 1942 and Enrico Fermi is reported to coin the backronym \textit{Safety Control Rod Axe Man}.} 
phase: Instead of a final explosion at an infinite temperature, the black hole turns into a light-weight remnant with vanishing temperature \cite{Nic09,DeS78,BaB88,Nic05,NSS06a,NiS10,BoR00,BoR06,APS01,IMN13,CMN15,%
Nic12,MMN11,Mod04,Mod06,DvG10,DvG12,DvG13b,FKN16,NiS12,COY14,NiS14,CMS14,SpS16,%
SpS16b,SpS15,XCal14}.

If the hole is hot enough, the evaporation produces all Standard Model particles on nearly equal footing since no gauge interactions are involved. Due to the flavor and color degrees of freedom and including antiparticles, there are 36 different quarks compared with 6 different charged leptons. Therefore, around 75\% of the primary emission are quarks and gluons. A detector, however, will detect a different composition. Because of mutual interactions of the emitted particles, plasmas around the black holes can arise, a chromosphere due to QCD reactions, and a QED-related photosphere \cite{CaS06}. Furthermore, the decay of instable particles will enhance the fraction of photons, neutrinos, electrons, and positrons. In experiments, data analysis concentrates on events with multiple high energetic objects. For semiclassical black holes, a lack of transverse energy is expected due to the emission of gravitons and neutrinos. In the case of quantum black holes, only few high-energy particles are supposed to be produced. The evaporation is expected to take place within $\SI{e-27}{\second}$ for semiclassical black holes and even shorter  for quantum black holes \cite{CMS2017}. Conversely it has been argued that the SCRAM phase would elongate the evaporation time  ($\sim \SI{e-16}{\second}$), due to the reduced emission in the final stage of the evaporation  \cite{CaN08,NiW11}.

\begin{figure*}[htbp]
\begin{center}
\begin{minipage}{0.45\textwidth}
	\includegraphics[width=\linewidth]{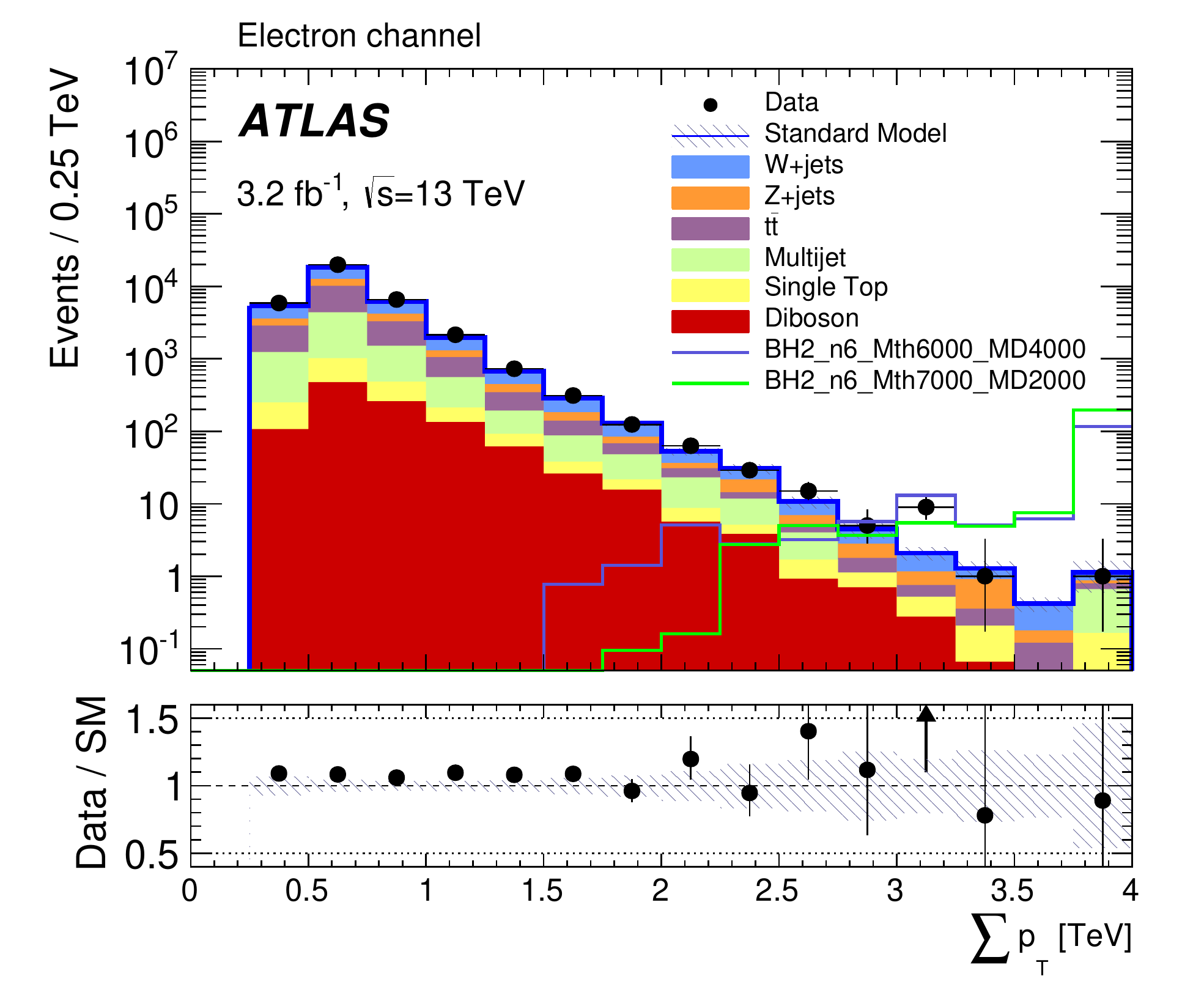}
	\label{fig:ATLAS_spectrum}
\end{minipage}
\hspace{0.05\textwidth}
\begin{minipage}{0.45\textwidth}
	\includegraphics[width=\linewidth]{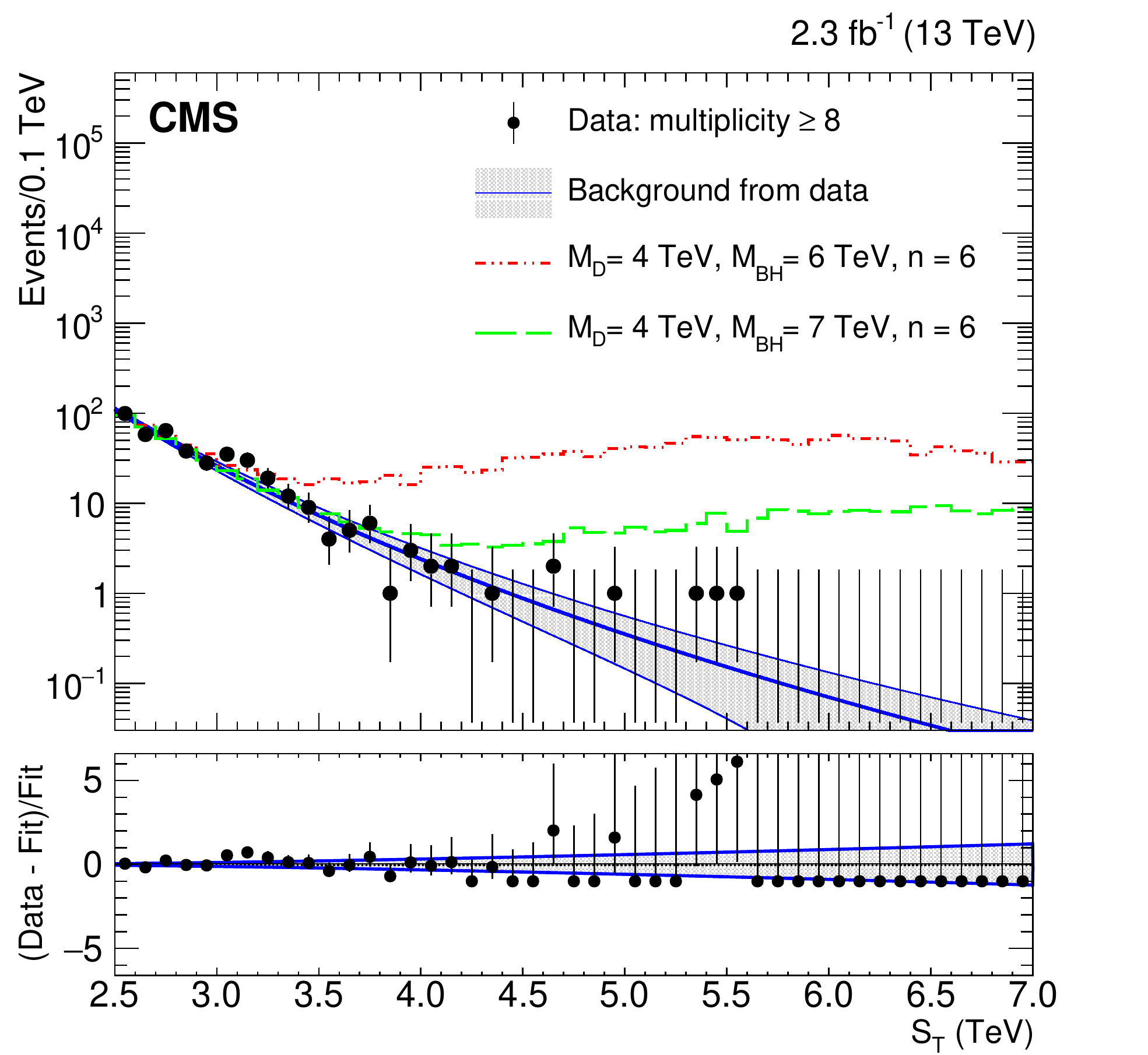}
	\label{fig:CMS_spectrum}
\end{minipage}
\caption{LHC Run 2 data at $\sqrt{s} = \SI{13}{\TeV}$ compared to Standard Model predictions. \newline
\textit{Left:} The dots and their error bars show the experimental spectrum of events with high transverse momenta, $p_\text{T}$, objects. Only those events are considered which posses a high-$p_\text{T}$ electron and at least 2 further objects with large $p_\text{T}$. The histogram indicates the theoretical contributions of different processes according to the Standard Model. The lower panel shows the ratio of the experimental counts to the Standard Model expectations.
The blue and green curves display black hole evaporation spectra simulated by the Charybdis 2 generator for rotating semiclassical black holes in 6 extra dimensions.\newline
\textit{Right:} An analogue graph based on the final particles' total transverse energy, $S_\text{T}$, for a combined jet, lepton, and photon multiplicity greater than 7. The lower panel shows the relative deviation of the data from the Standard Model predictions. The red and the green curve provide the simulated evaporation spectra of two semiclassical black holes for reference.
	Figures from \cite{ATLAS2016,CMS2017}.}
\label{fig:Spectra}
\end{center}
\end{figure*}

\begin{figure*}[htbp]
\begin{center}
	\includegraphics[width=0.45\textwidth]{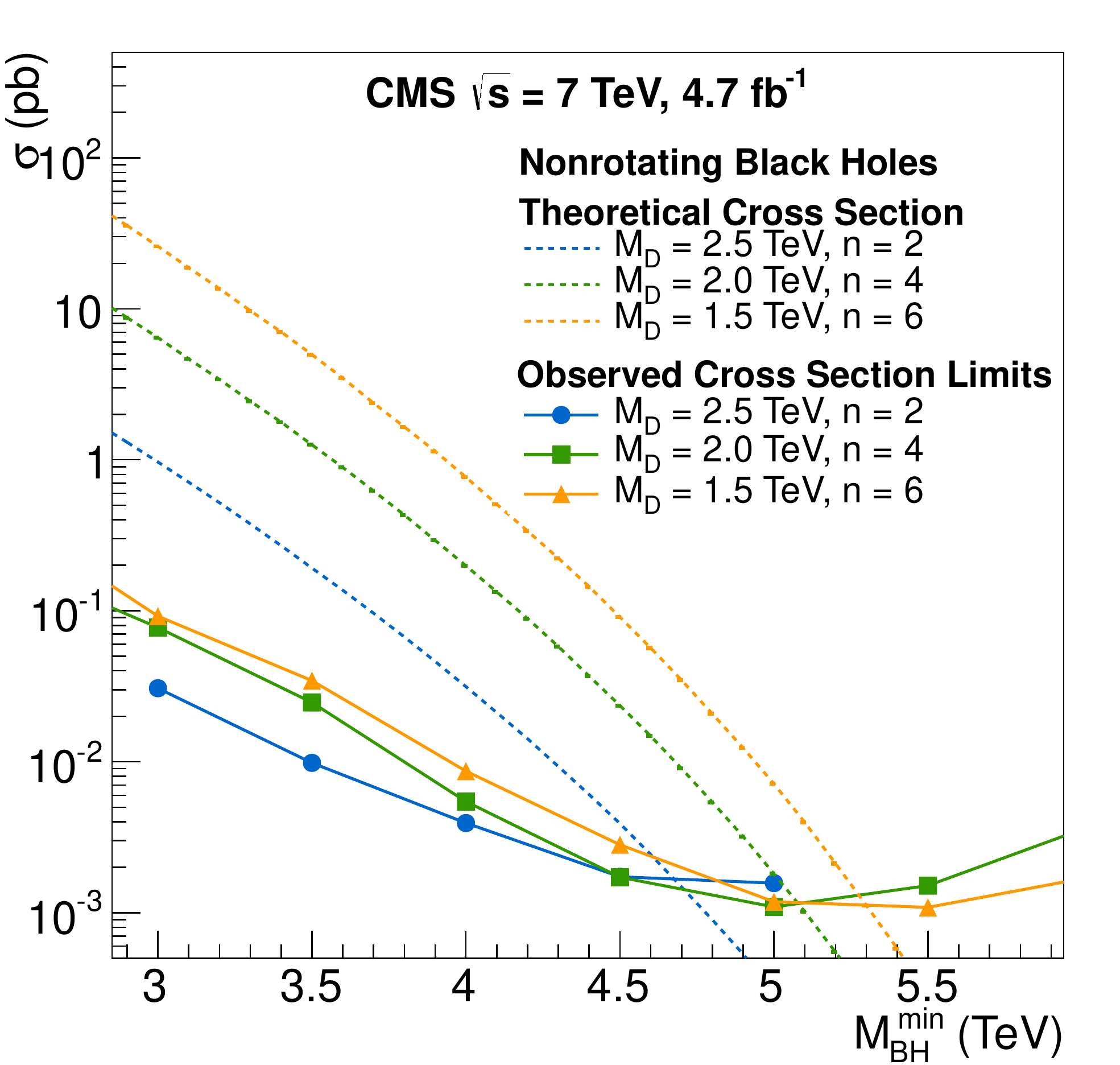}
\caption{Experimental limits on the black hole production cross section as a function of the minimal black hole mass, $M^\mathrm{min}_\mathrm{BH}$, from LHC Run 1 at $\sqrt{s} = \SI{7}{\TeV}$. The solid lines provide the experimentally derived upper bounds for a given fundamental energy scale, $M_D$, and number of extra dimensions, $n$. The dotted curves show the expected cross sections simulated by the BlackMax generator for complete evaporation. $M^\mathrm{min}_\mathrm{BH}$ is constrained to be larger than approx.~$\SI{5}{\TeV}$.
	Figure from \cite{CMS2012}.}
\label{fig:Cross_Sections}
\end{center}
\end{figure*}

\begin{figure*}[htbp]
\begin{center}
\begin{minipage}{0.45\textwidth}
	\includegraphics[width=\linewidth]{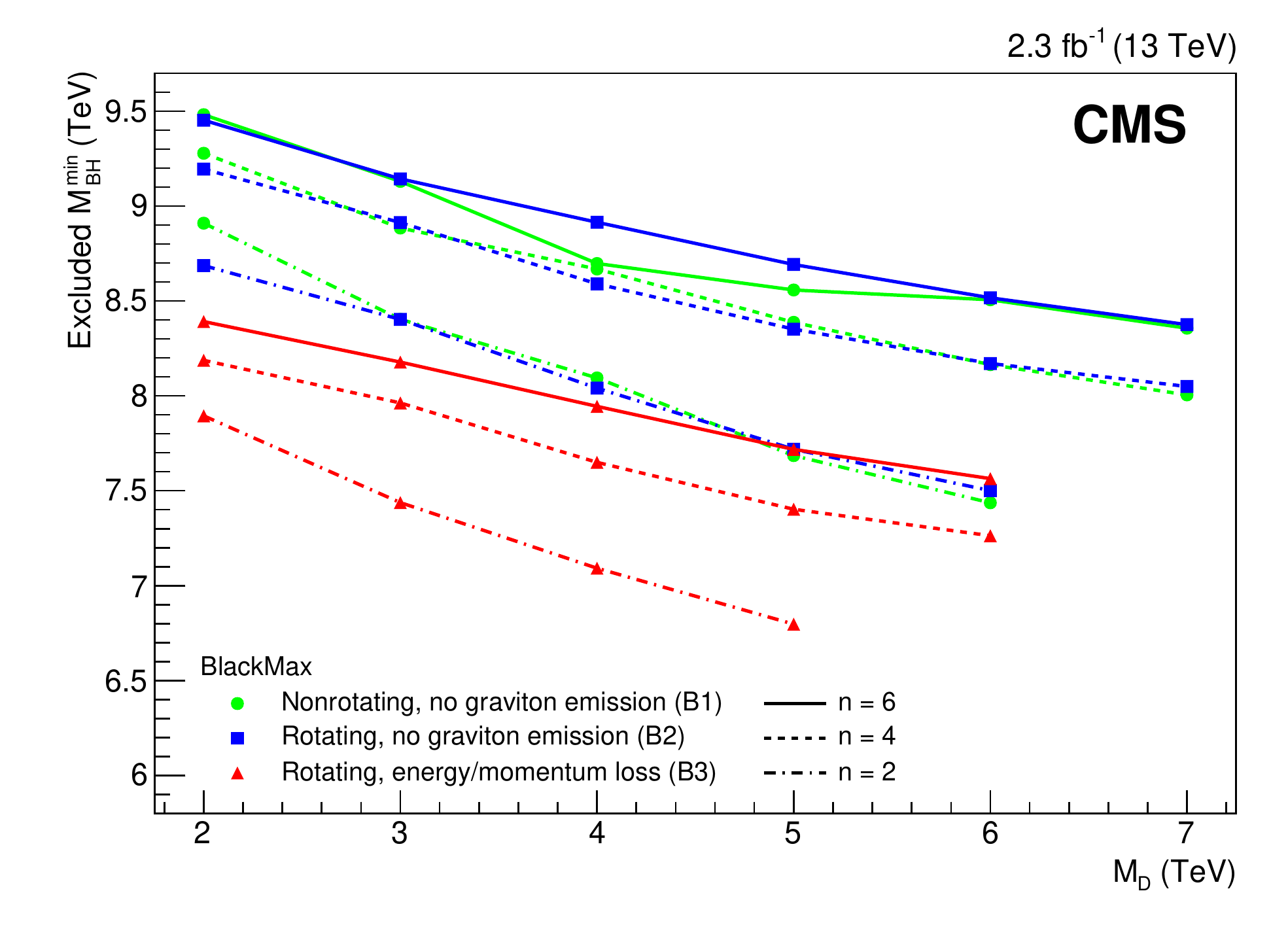}
	\label{fig:CMS_lower_bounds_Mmin_BlackMax}
\end{minipage}
\hspace{0.05\textwidth}
\begin{minipage}{0.45\textwidth}
	\includegraphics[width=\linewidth]{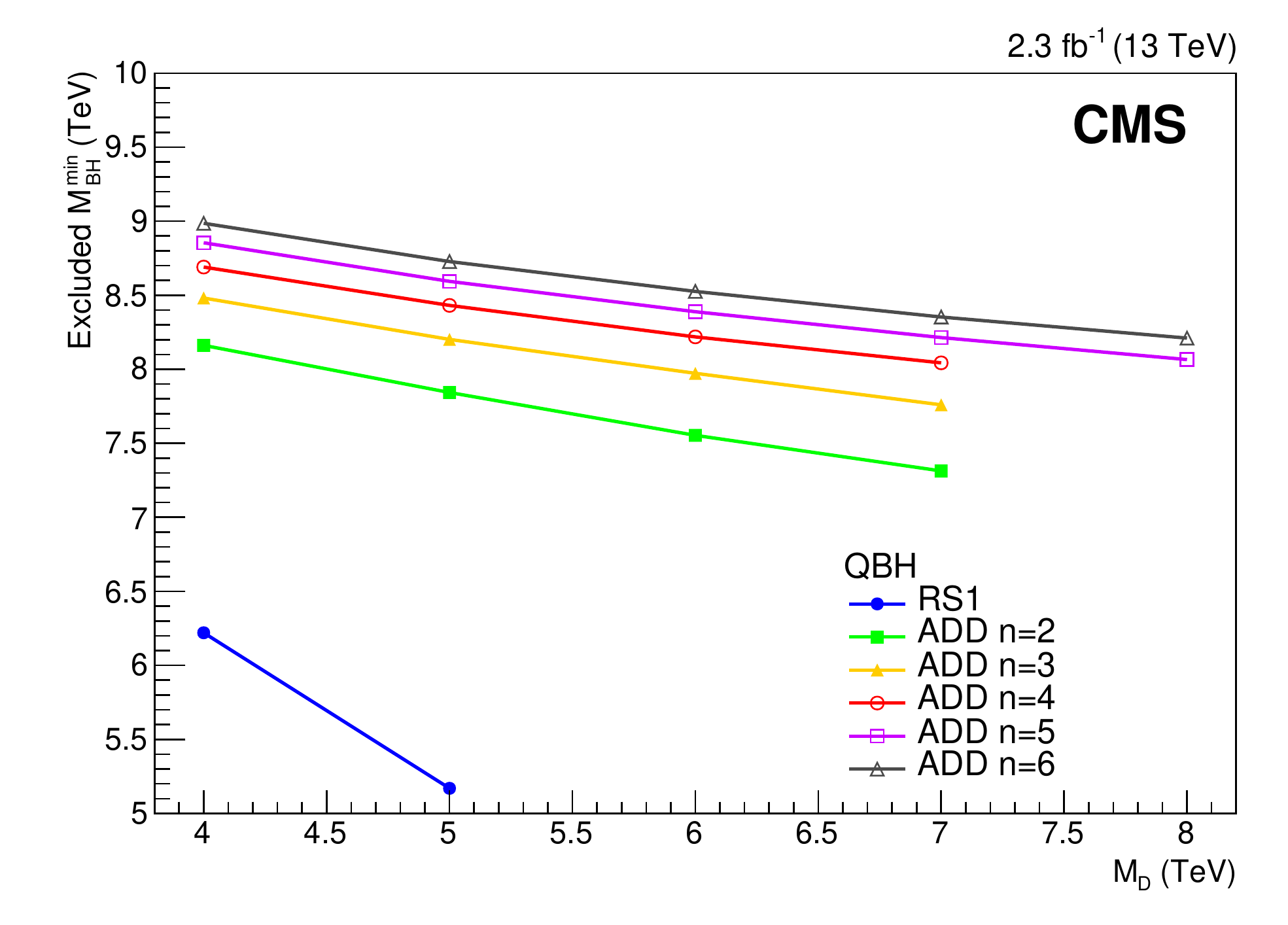}
	\label{fig:CMS_lower_bounds_Mmin_QBH}
\end{minipage}
\caption{Lower threshold for the minimal black hole mass, $M^\mathrm{min}_\mathrm{BH}$, from LHC Run 2 at $\sqrt{s} = \SI{13}{\TeV}$ as a function of the fundamental gravitational energy scale, $M_D$, for different numbers of extra dimensions, $n$.\newline
\textit{Left:} Constraints for semiclassical black holes simulated with the BlackMax generator.\newline
\textit{Right:} Constraints for quantum black holes simulated with the QBH generator. The limit for a Randall-Sundrum extra dimensional model (RS1) is indicated for comparison.
	Figures from \cite{CMS2017}.}
\label{fig:Mmin}
\end{center}
\end{figure*}

At the LHC, the ATLAS and CMS collaborations look for black hole signatures. Up to now, no hint for black holes has been found and there is no significant deviation from the expected Standard Model background, cf.~Fig.~\ref{fig:Spectra}. Theoretical predictions for the black hole production cross sections are obtained from event generators: BlackMax \cite{DaiSSIRT2008} and Charybdis 2 \cite{HarrisRW2003,HarrisK2003,FrostGSCDPW2009} for semiclassical black holes, and QBH \cite{Gingrich2010a,Gingrich2010b} for a model of quantum black holes. Comparison of these theoretically predicted with experimental upper limits for the cross sections leads to lower bounds on the minimal black hole mass, $M^\mathrm{min}_\mathrm{BH}$. This is illustrated in Fig.~\ref{fig:Cross_Sections} for a CMS analysis of LHC Run 1 with $\sqrt{s} = \SI{7}{\TeV}$. The latest CMS analysis from Run 2 at $\sqrt{s} = \SI{13}{\TeV}$ \cite{CMS2017} has increased the constraints on the minimal black hole mass. For semiclassical black holes and for quantum black holes, $M^\mathrm{min}_\mathrm{BH}$ has to be to larger than $\SI{7}{\TeV}$ to $\SI{9}{\TeV}$, cf.~Fig.~\ref{fig:Mmin}. This is consistent with the constraints derived by ATLAS \cite{ATLAS2016}.

\section{Conclusions and Outlook}
\label{sec:over&out}
Recent experimental progress has started an exciting time for gravitational physics and astrophysics. The time of direct observations of black holes seems to become reality. Black holes in the microscopic, stellar, and supermassive regime can be targeted by dedicated improved investigation methods. 

Theoretically, gravitational wave signals and black hole shadows can be predicted by most alternative gravitational theories. Experimentally, they can be recorded by corresponding observatories like LIGO and the EHT. By comparison, Einstein's general theory of relativity undergoes strong-field tests against alternative theories like pseudo-complex general relativity. Among possible classical deviations, maybe quantum gravity effects manifest themselves in some hidden details.

As a possible solution to the hierarchy problem, the ADD model proposes compactified large extra dimensions which are only accessible for gravitational interactions. Consequently, the fundamental gravitational energy scale, $M_D$, is drastically decreased; it could lie in the range of present particle accelerators. The LHC reaches energies which might allow the production of black holes. Their traces due to evaporation, namely additional events with a high multiplicity of large transverse momentum objects, have not yet been detected. Therefore the lower limit for the minimal black hole mass has been increased to $\SI{7}{\TeV}$ to $\SI{9}{\TeV}$ in LHC Run 2.

Further developments rely on improvements of the experiments. By decreasing the noise in gravitational wave observatories, neutron star mergers should come into reach. The gravitational wave signal is supposed to constrain the neutron star equation of state and thus the QCD one, too. A higher resolution of the EHT will bring up more details of accretion physics and jet launching. Accordingly,  the tests on Einstein's general relativity will become tighter. 

On the microscopic black hole side, black hole signatures might be uncovered in the next LHC Runs due to the expected increase of beam luminosity. We expect data to allow for conclusions about final stages of the black hole evolution, namely the quantum phase where semiclassical gravity breaks down. In particular, it would be a great step towards quantum gravity if there were evidence for a SCRAM phase.

\section*{Acknowledgment}
Our interest in black hole physics was strongly motivated by our teacher and friend Walter Greiner. We dedicated this article to his memory. In addition to Walter's guidance and support further support was provided by the Frankfurter F\"orderverein f\"ur physikalische Grundlagenforschung, by the Stiftung Polytechnische Gesellschaft Frankfurt am Main, and DFG.


\providecommand{\href}[2]{#2}\begingroup\raggedright\endgroup

\end{document}